\documentclass[final,5p,times,twocolumn]{elsarticle}

\usepackage{lineno,hyperref}
\modulolinenumbers[5]

\usepackage{graphicx}
\usepackage{amssymb}
\usepackage{wasysym}
\journal{Nuclear Instruments and Methods A}

\begin{document}

\begin{frontmatter}

\title{First demonstration of VUV-photon detection in liquid xenon with THGEM and GEM-based Liquid Hole Multipliers}

\author{E. Erdal}

\author{L. Arazi\fnref{mycorrespondingauthor}}
\fntext[mycorrespondingauthor]{Corresponding author}

\ead{lior.arazi@weizmann.ac.il}

\author{M. Rappaport, S. Shchemelinin, D. Vartsky, and A. Breskin}

\address{Department of Particle Physics and Astrophysics, Weizmann Institute of Science, Rehovot 7610001, Israel}

\begin{abstract}
The bubble-assisted Liquid Hole-Multiplier (LHM) is a recently-introduced detection concept for noble-liquid time projection chambers. In this ``local dual-phase'' detection element, a gas bubble is supported underneath a perforated electrode (e.g., Thick Gas Electron Multiplier - THGEM, or Gas Electron Multiplier - GEM). Electrons drifting through the holes induce large electroluminescence signals as they pass into the bubble. In this work we report on recent results of THGEM and GEM electrodes coated with cesium iodide and immersed in liquid xenon, allowing \textemdash for the first time \textemdash the detection of primary VUV scintillation photons in addition to ionization electrons. 
\end{abstract}

\begin{keyword}
Noble-liquid detectors\sep Liquid Hole-Multipliers\sep Micro-Pattern Gas Detectors (MPGD)\sep Gas Electron Multiplier (GEM) \sep Thick Gas Electron Multiplier (THGEM) \sep cesium iodide
\MSC[2010] 00-01\sep  99-00
\end{keyword}

\end{frontmatter}


\section{Introduction}
Liquid Hole-Multipliers (LHMs) were recently proposed as a new detection concept of both VUV-photons and ionization electrons induced by particle interaction in noble liquids~\cite{Breskin_LHM0}. The original motivation was to find a solution to the challenge of maintaining a uniform electroluminescence (EL) response across the large diameter of future multi-ton dual-phase noble-liquid time projection chambers (TPCs) for dark matter detection; however, the concept may also be applicable in other fields, including neutrino physics, Compton imaging, and neutron and gamma radiography. In its basic conceptual form, an LHM consists of a perforated electrode, e.g., a Thick Gas Electron Multiplier (THGEM)~\cite{Breskin_THGEM_review} or a Gas Electron Multiplier (GEM)~\cite{Sauli_GEM}, immersed in a noble liquid; a VUV-sensitive cesium iodide (CsI) photocathode is optionally deposited on the electrode top surface. Radiation-induced ionization electrons and photoelectrons ejected from the photocathode are focused by an applied electric field into the electrode's holes where they create EL signals. The expectation that LHMs could be used for VUV photon detection stemmed from a previous demonstration~\cite{Aprile_Peskov_CsI} of the efficient extraction of photoelectrons from CsI into LXe, with a quantum efficiency (QE) reaching $\sim$30\% at 175~nm for a field of 20~kV/cm.\par
Preliminary measurements with a THGEM electrode immersed in liquid xenon (LXe)~\cite{Arazi_LHM1} showed a large EL yield at relatively low voltages. Later studies demonstrated that the process in fact occurs within a xenon gas bubble trapped under the electrode surface~\cite{Arazi_bubbles,Erdal_bubbles}. This so-called bubble-assisted EL mechanism was found to be stable over months of operation; under specific conditions it was shown to yield up to 7.5\% RMS energy resolution for ionization electrons from 5.5~MeV alpha particles stopped inside the liquid~\cite{Erdal_bubbles}. The estimated light yield was a few dozen photons per electron (emitted into 4$\pi$) at a THGEM voltage $\Delta V_{THGEM}\approx3000$~V~\cite{Arazi_bubbles}. \par
In this work we extend the study to the detection of VUV-photons, in addition to ionization electrons, for both THGEM and GEM electrodes coated with CsI and immersed in LXe using alpha particles. The motivation for GEM-based LHMs is two-fold: first, the electric field on the surface of the photocathode is much higher for a GEM than for a THGEM, and thus can lead to higher photoelectron extraction efficiency into the liquid; second, the GEM's smaller dimensions can reduce the spread of travel times of photoelectrons emitted from CsI at varying distances from the holes, thus providing better time resolution. We begin by demonstrating the bubble-assisted EL process in a bare GEM (with no CsI). We then demonstrate, for the first time, the detection of primary scintillation photons with CsI-coated THGEM and GEM electrodes, providing information on both the energy and time resolution of the process. 

\section{Experimental setup}
All measurements were carried out in a LXe cryostat, described in detail in~\cite{Erdal_bubbles}. 
A grating of parallel resistance wires below the THGEM/GEM electrode was used for controlled bubble formation by ohmic heating. An external camera allowed for direct observation of the bubble formed at the bottom of the THGEM/GEM. The setup (figure 1) comprised a $\sim$180~Bq spectroscopic $^{241}$Am source emitting 5.49~MeV alpha particles into the liquid, located $\sim$6 or 10~mm above the THGEM/GEM electrode, which was either bare or coated with CsI on its top surface. For the VUV-photon detection measurements (section 3.2), an 85\% transparent mesh was placed 5~mm below the source to create a uniform drift field ($E_{drift}$) above the THGEM/GEM. The wire grating was placed 4.5~mm below the electrode; in the present experiments one central wire was used for heating, with the remaining wires (at ground potential) defining the transfer field ($E_{transfer}$). The $^{241}$Am source holder, mesh, THGEM/GEM top and bottom faces, and wire grating were separately biased to produce the desired drift, THGEM/GEM, and transfer fields. Two 1'' square PMTs (Hamamatsu R8520) were used: the top one for triggering on alpha-particle S1 primary scintillation (by photon reflections from the surrounding PTFE structure), and the bottom one for recording the EL S2 and S1' (see below) light. The PMT signals were digitally recorded on an oscilloscope (Tektronix MSO5204B) and later processed by dedicated software. The liquid level was kept above the top PMT face. Xenon was circulated throughout the experiments at 2~slpm through an SAES noble-gas purifier.

\begin{figure}[hbt] 
	\centering 
	\includegraphics[width=\columnwidth,keepaspectratio]{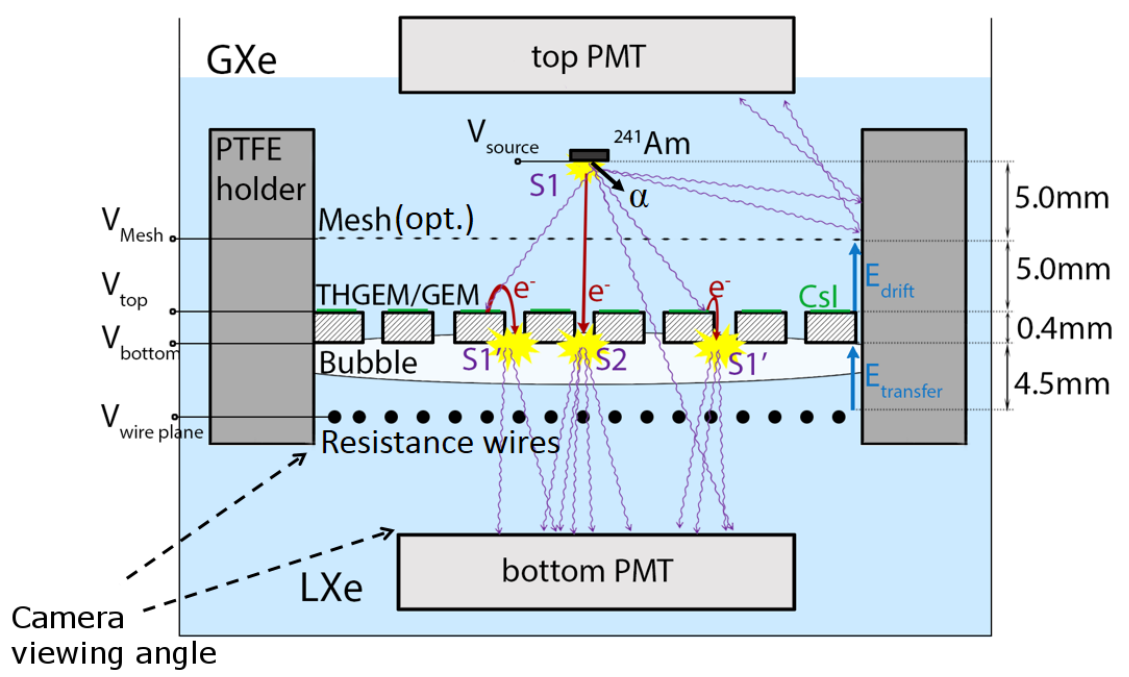}
	\caption{Schematic view of the experimental setup for the recording of photoelectron and ionization electron signals from a THGEM/GEM electrode immersed in LXe, comprising an $^{241}$Am alpha-particle source, mesh (optional), THGEM/GEM and resistance wire grating for generating bubbles. Signals are recorded from the bottom PMT, triggered on alpha-particle primary scintillation signals from the top PMT. An external camera views the THGEM/GEM from below. When the mesh is not present $E_{drift}$ is defined as the field between the THGEM/GEM top and the source (now $\sim6$~mm above the former.)}
\end{figure}

The circular THGEM electrode had the same geometry as in~\cite{Arazi_LHM1,Arazi_bubbles,Erdal_bubbles}, namely: 0.4~mm-thick FR4 substrate with Au-coated Cu clad on both sides; an active area 34~mm in diameter; an hexagonal array of $\diameter$0.3~mm holes with 1~mm pitch and 0.1~mm-wide etched rims. The GEM electrodes had an active 14~mm-diameter central region of ``standard geometry'': 50~$\mu$m-thick Kapton with Au-coated 5~$\mu$m-thick Cu on both sides;  double-conical holes ($d_{out}=70$~$\mu$m, $d_{in}=50$~$\mu$m); hexagonal lattice of 140~$\mu$m pitch. CsI photocathodes were vacuum-deposited on the THGEM/GEM top faces following standard procedures~\cite{Breskin_CsI_review}. The THGEM was baked at 100$^{\circ}$C for 2~hours in dry N$_2$ prior to evaporation; the GEM electrodes were not baked out.
The nominal thickness of the photocathodes was 300~nm; their QE was estimated \textit{in situ} inside the evaporation chamber in vacuum using a mercury lamp, with a typical value of $\sim$18\% at 185~nm. Assembly of the detector setup was done in a glove box continuously flushed with dry N$_2$. During installation the cryostat was flushed with Ar to avoid exposure of the CsI layer to humidity.

\section{Results}

\paragraph{3.1.	Electroluminescence in the holes of a GEM} Similarly to previous observations with immersed THGEM electrodes~\cite{Erdal_bubbles}, a large bubble was formed below the GEM either spontaneously or by running $\sim$0.7~A through the resistive wire for $\sim$10~s, remaining stable thereafter. In steady state the bubble entirely filled the area below the GEM (limited by the internal wall of the PTFE holder), with a thickness of $\sim$3-4~mm (figure~1). Figure~2 shows a photograph taken during the initial growth of the bubble, partly covering the GEM bottom surface.

\begin{figure}[hbt] 
	\centering 
	\includegraphics[width=0.4\textwidth,keepaspectratio]{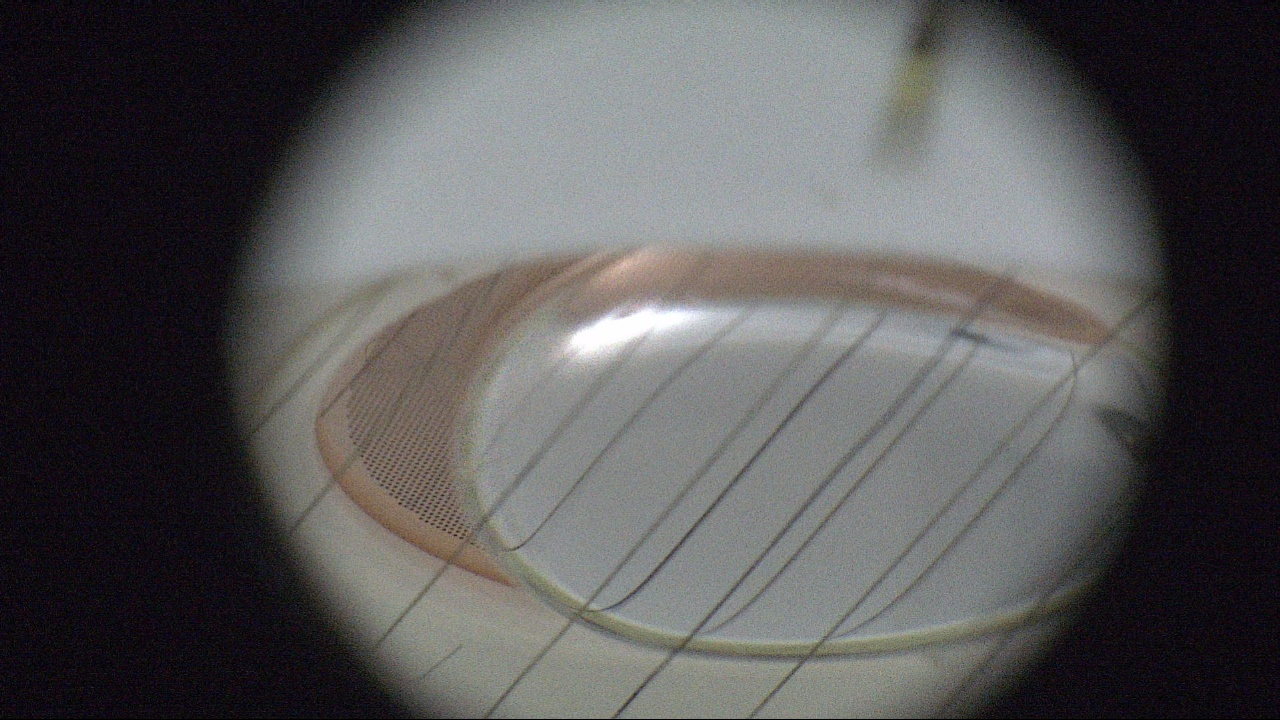}
	\caption{Bubble sustained under a GEM electrode in LXe. The photograph was taken during the initial growth of the bubble; in steady state it covers the full area below the electrode.}
\end{figure}

Initial EL measurements were done with a bare GEM (with no CsI) and no mesh (the source, in this case, was $\sim$6~mm above the GEM). Alpha-particle-induced ionization electrons were collected from the source under a $\sim$0.1-1~kV/cm drift field. $E_{transfer}$ was either zero, $-$2.4~kV/cm (directing the electrons passing through the holes to the GEM bottom), or $+$0.67~kV/cm (transporting the electrons towards the wires). As in previous works~\cite{Arazi_bubbles,Erdal_bubbles} EL S2 signals were observed only in the presence of a bubble; with no bubble present, no S2 signals were detected for any GEM voltage ($\Delta V_{GEM}$) up to the onset of discharges at $\Delta V_{GEM}=1600$~V, corresponding to a maximum field of $\sim$200~kV/cm at the center of the GEM holes. (see~\cite{Erdal_bubbles} for details on the procedures for bubble formation and elimination.) With a bubble present in steady state, S2 signals appeared at $\Delta V_{GEM}\sim200$ V, increasing linearly with the voltage. Figure~3 shows an example of a waveform recorded from the bottom PMT (biased at $-600$~V) for $\Delta V_{GEM}=1200$~V, with both S1 and S2 signals present (S1 originates from the small fraction of primary scintillation photons emitted from the alpha-particle track, that reach the bottom PMT through the GEM holes). The typical width of the S2 pulses was $\sim$150~ns FWHM, compared to $\sim$300~ns with an immersed THGEM~\cite{Arazi_bubbles}.

\begin{figure}[hbt] 
	\centering 
	\includegraphics[width=\columnwidth,keepaspectratio]{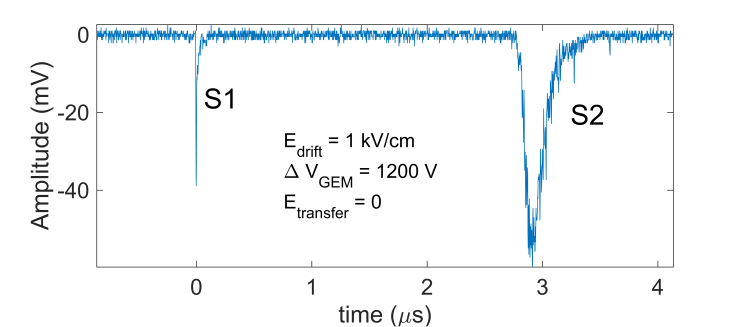}
	\caption{Typical single-event S1 and S2 signals from a bare GEM (with no CsI) in LXe. Setup of figure 1, without the mesh; the source is located $\sim6$~mm above the GEM.}
\end{figure}

Figure~4 shows the magnitude (integrated bottom-PMT pulse area) as well as the RMS resolution of the GEM S2 signal compared to that of a THGEM (taken from~\cite{Erdal_bubbles}). The signal depends linearly on $\Delta V_{THGEM/GEM}$ (and hence on the electric field) as expected from an EL process. Interestingly, the signal magnitude is similar for the same applied voltage on both electrodes. The significantly better resolution at lower voltages in the GEM is attributed to better electron collection into its holes. $E_{drift}$ was 0.88~kV/cm and 0.36~kV/cm for the GEM and THGEM, respectively.

\begin{figure}[hbt] 
	\centering 
	\includegraphics[width=\columnwidth,keepaspectratio]{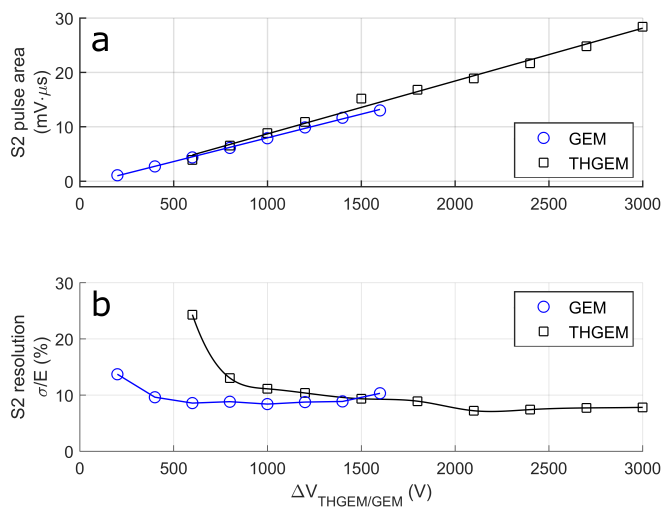}
	\caption{S2 magnitude (a) and RMS resolution (b) as a function of $\Delta V_{THGEM/GEM}$ for alpha particles (setup of figure 1).}
\end{figure}

\paragraph{3.2. VUV-photon detection with CsI-coated THGEM and GEM} VUV-photon detection was demonstrated for both immersed THGEM and GEM electrodes coated with CsI (setup of figure~1). In this case, the observed waveforms (figure~5) contained an additional EL signal (here termed S1') immediately following the primary scintillation pulse (S1). S1' results from the emission of photoelectrons from the CsI photocathode by alpha-particle-induced primary scintillation photons; these are focused by the dipole-like field into the holes, inducing EL in the underlying bubble, similarly to the mechanism governing S2 signal formation. While the S1' signal in a THGEM lags S1 by $\sim250$~ns (figure~5a), that of the GEM arrives $\lesssim30$~ns after S1, appearing during the decay of the primary scintillation signal (figure~5b). This is attributed to the shorter distance between the photoelectron starting point and the bubble top surface in the thinner, smaller-pitch GEM structure. The ratio between the measured S1' and S2 signals is considerably larger here for the GEM compared to the THGEM, implying a higher overall detection efficiency of the VUV photons. This is qualitatively expected based on the higher field at the CsI-coated GEM surface. A quantitative study of S1' in GEMs and THGEMs will be given in a separate publication.

\begin{figure}[hbt] 
	\centering 
	\includegraphics[width=\columnwidth,keepaspectratio]{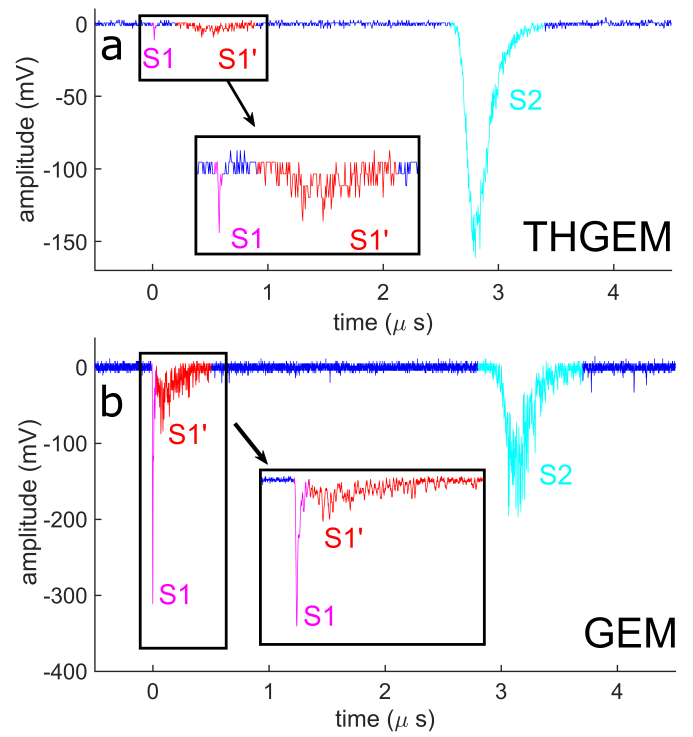}
	\caption{Typical single-event S1-S1'-S2 waveforms induced by alpha particles, recorded by the bottom PMT for a THGEM (a) and GEM (b) (setup of figue 1).}
\end{figure}

The magnitude of the S1' signal and its energy and time resolutions are shown for a CsI-coated THGEM in figure~6, as a function of $\Delta V_{THGEM}$ for various $E_{drift}$ and $E_{transfer}$ values. The signal magnitude is growing faster than linear as a function of $\Delta V_{THGEM}$; we attribute this effect to an increase in the extraction efficiency of photoelectrons from the photocathode for higher $\Delta V_{THGEM}$. When applying $E_{drift}=1.2$~kV/cm, pulling ionization electrons towards the THGEM, the magnitude of S1' is smaller because of the reduced net field on the photocathode surface. The direction of the transfer field between the THGEM and wire grating (determining whether the electron trajectories terminate on the THGEM bottom or the grating) has no effect on S1', as already demonstrated in~\cite{Arazi_bubbles}. Qualitatively similar behavior of S1' was observed with a CsI-coated GEM as a function of $\Delta V_{GEM}$. While the RMS energy resolution was roughly the same as that of the THGEM S1', the GEM time resolution was considerably better: $\sim4$~ns RMS for several hundred photoelectrons. Figure~7 shows the S1' magnitude and time distributions (measured with respect to S1) for a GEM operated at 1200~V and 1300~V, respectively, in this case with $E_{drift}=0$. S1' timing was defined in post-processing as the point for which the integrated pulse area reached 30\% of its total value.

\begin{figure}[hbt] 
	\centering 
	\includegraphics[width=\columnwidth,keepaspectratio]{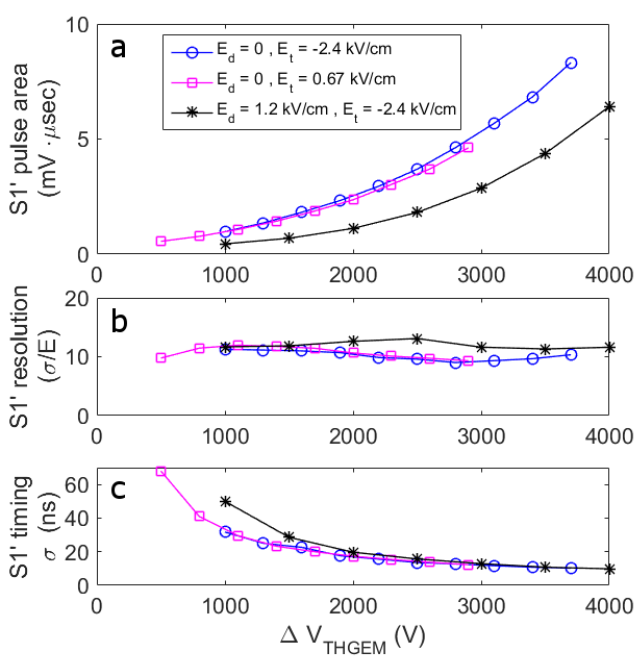}
	\caption{THGEM: S1' magnitude, energy- and time resolutions for alpha particle-induced signals, as function of $\Delta V_{THGEM}$ (setup of figure 1).}
\end{figure}

\begin{figure}[hbt] 
	\centering 
	\includegraphics[width=0.45\textwidth,keepaspectratio]{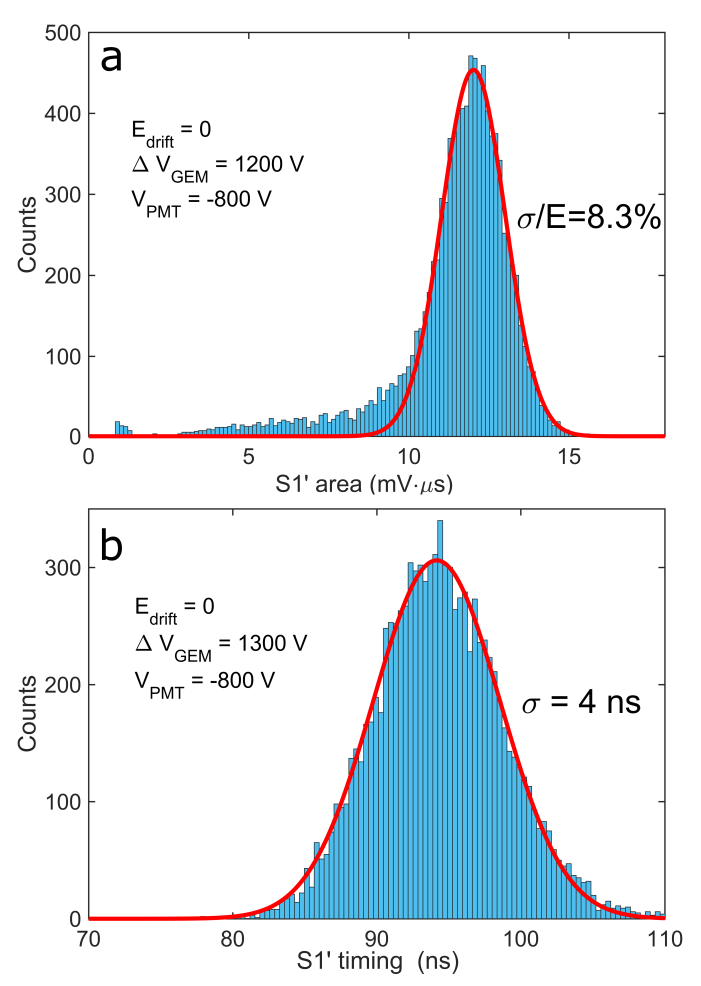}
	\caption{GEM: S1' energy (a) and timing (b) distributions for alpha-particle-induced signals (setup of figure 1).}
\end{figure}

\section{Conclusion}
The present study expanded our previous work on the bubble-assisted LHM concept using THGEMs to GEM electrodes immersed in LXe. It demonstrated, for the first time, the ability of CsI-coated LHM elements to detect both primary VUV scintillation photons and ionization electrons, under the same operation conditions. Similarly to the THGEM, the EL process with immersed GEM electrodes (bare and CsI-coated) relies on the presence of a trapped xenon gas bubble underneath. The VUV-induced EL signals yielded a pulse-height resolution of $\sim8\%$ RMS for a few hundred photoelectrons. Although inherently more delicate, the advantages of GEM LHMs over THGEM LHMs are their higher surface field \textemdash leading to a higher expected photon detection efficiency \textemdash and better timing properties, due to the more compact hole pattern. Ongoing studies focus on the systematic characterization of CsI-coated LHMs (GEMs and THGEMs), with a particular emphasis on maximizing their photon detection efficiency and EL light yield. While a primary motivation is the potential use of bubble-assisted CsI-coated LHM modules in future noble-liquid dark matter TPCs, LHMs might also be attractive in other types of searches and applications utilizing noble liquid detectors.   

\section{Acknowledgments}
The authors thank Prof. V. Chepel of Coimbra University for useful discussions. This work was partly supported by the Minerva Foundation with funding from the German Ministry for Education and Research (Grant No. 712025) and the Israel Science Foundation (Grant No. 791/15). The research was carried out within the DARWIN consortium for a future multi-ton LXe dark matter observatory. A. Breskin is the W.P. Reuther Professor of Research in The Peaceful Use of Atomic Energy.


\bibliography{mybibfile}

\end{document}